# Directional Connectivity-based Segmentation of Medical Images

Ziyun Yang
Duke University
Durham, NC, United States
ziyun.yang@duke.edu

Sina Farsiu
Duke University
Durham, NC, United States
sina.farsiu@duke.edu

## Abstract

*Anatomical consistency in biomarker segmentation is crucial for many medical image analysis tasks. A promising paradigm for achieving anatomically consistent segmentation via deep networks is incorporating pixel connectivity, a basic concept in digital topology, to model inter-pixel relationships. However, previous works on connectivity modeling have ignored the rich channel-wise directional information in the latent space. In this work, we demonstrate that effective disentanglement of directional sub-space from the shared latent space can significantly enhance the feature representation in the connectivity-based network. To this end, we propose a directional connectivity modeling scheme for segmentation that decouples, tracks, and utilizes the directional information across the network. Experiments on various public medical image segmentation benchmarks show the effectiveness of our model as compared to the state-of-the-art methods. Code is available at https://github.com/Zyun-Y/DconnNet.*

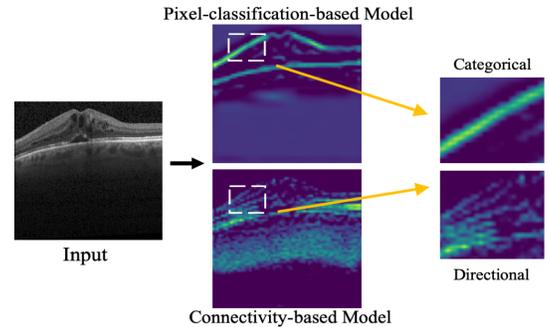

Figure 1. The latent space differences between traditional pixel-classification-based and connectivity-based models. In the former, only categorical features, e.g., boundaries, are highlighted; while in the latter, the feature map also contains directional information (e.g., the horizontal connections between boundary pixels).

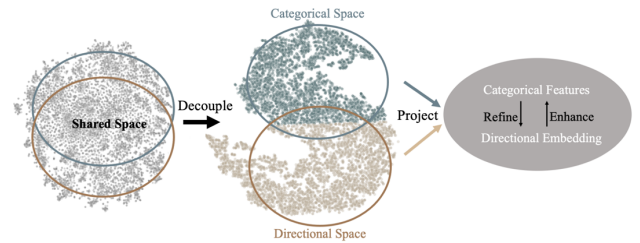

Figure 2. The flows of the two groups of latent features (categorical and directional) in the latent space of DconnNet, are visualized by T-SNE [13]. They were first disentangled (Sec 3.2) and then effectively fused in a projected shared manifold (Sec 3.3). The colors are rendered based on the results of clustering.

## 1. Introduction

Maintaining anatomical consistency in the segmentation of medical images is important but challenging, as minor geometric errors may change the global topology [1, 2] and cause functional mistakes in downstream clinical decision-making [3]. Anatomical consistency in images can be expressed with topological properties, such as pixel connectivity and adjacency [4, 5]. As such, by directly modeling the mutual information between pixels or regions, graph-based methods have long been used to correct topological and geometrical errors [6-8]. However, such classic machine vision techniques usually depend on manually defined priors and thus are not easily generalizable for a wide variety of applications.

Alternative to the classic approaches, deep learning-based segmentation methods utilized an encoder-decoder architecture [9] to learn from a group of pixels in a particular receptive field at each layer. More recently, significant progress has been made in capturing the inter-pixel dependency inside a network's latent space [10-12]; however, very few studies have been conducted on the problem modeling side of the networks. A typical segmentation network models the problem as a pure pixel-wise classification task and uses a segmentation mask as the only label. Yet, this pixel-wise modeling scheme is suboptimal as it does not directly exploit inter-pixel relationships and geometrical properties [14, 15]. Thus, these models may result in low spatial coherence (i.e., inconsistent predictions for neighboring pixels that share similar spatial features) in their prediction [16]. Especially, when applied to high noise/artifacts medical data, the lower spatial consistency may lead to topological issues [17].

The concept of pixel connectivity has long been used to ensure the basic topological duality of separation and connectedness in digital images [18]. More recently, in the context of deep learning, the connectivity masks, reviewed in Section 2.1, have been introduced as a topological extension of the segmentation mask [15]. Using connectivity masks as training labels has several advantages over segmentation masks. In terms of problem modeling, using a connectivity mask inherently changes the problem from pixel-wise classification to connectivity prediction, which models and enhances the topological representation between pixels of interest. In terms of label representation, a connectivity mask is more informative in three ways: first, a connectivity mask stores the categorical information among the connections of pixels and it is inter-pixel relation-aware; second, it sparsely represents edge pixels [16]; third, it contains rich directional information channel-wisely. Thus, a network trained with connectivity masks has both categorical (reflected by connectivity) and directional features in its latent space, each of which forms a specific sub-latent space, as shown in Fig. 1.

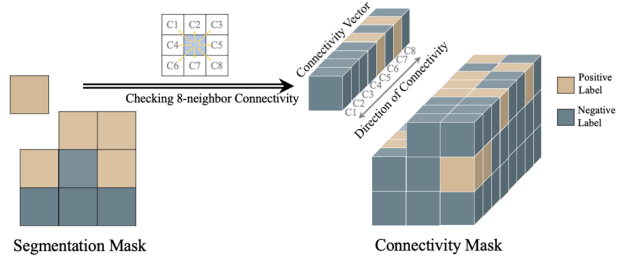

Figure 3. Illustration of generating the connectivity mask from the segmentation mask by traversing 8-neighbor pixel connectivity. For each pixel, channel $C_i$'s binary value in the converted connectivity vector carries the categorical information (connected or not connected) while $i$ encodes the directional information (direction of connection).

In previous studies [15, 16, 19-22], these two groups of features were learned simultaneously through a shared network path which may result in highly coupled latent space and introduce redundancy [23]. Further, effectively disentangling meaningful subspaces from the shared latent space has been shown effective in accounting for the dependencies/independencies between features [24, 25].

Inspired by the idea of latent space disentanglement, in this paper, we propose a novel directional connectivity-based segmentation network (DconnNet) to disentangle the directional subspace from the shared latent space and utilize the extracted directional features to enhance the overall data representation, as in Fig. 2. The disentangling process is conducted by a sub-path slicing-based module called Sub-path Direction Excitation (SDE). The directional-based feature enhancement is applied in a coarse-to-fine manner using an Interactive Feature-space Decoder (IFD) with two top-down interactive decoding flows. Finally, we propose a novel Size Density loss (SDL) that alleviates the common data imbalance problem in medical datasets with a label size distribution-based weighting scheme. With experiments on different public medical image analysis benchmarks, we demonstrate the superiority of DconnNet against other state-of-art methods.

## 2. Related Work
### 2.1. Deep connectivity modeling

In topology, pixel connectivity describes how neighboring pixels are related to each other [4]. Following broad utilization in characterizing topological properties in classic image processing methods [26, 27], connectivity has found new applications in deep learning-based image segmentation [15, 16, 19-22]. The connectivity-based segmentation networks use the connectivity mask (Fig. 3) as the label, defined as an 8-channel mask with each channel representing if a pixel on the original image belongs to the same class of interest with one of its neighboring pixels at a specific direction. The connectivity mask was first introduced and applied to image segmentation in [15]. This idea was later extended by other works, including [16] which showed the bilateral property of pixel connectivity in saliency detection, and [28] which modeled the cross-modality connectivity for radar-video data fusion. Meanwhile, the effectiveness of connectivity modeling has been demonstrated in different applications such as remote sensing segmentation [19], path planning [20], and medical image segmentation [21, 22, 29]. Although significant progress has been made in this field, we demonstrate that the rich directional information in the connectivity masks has not yet been fully utilized.

### 2.2. Interpretation of latent space

Latent space is an embedding of a set of features in a deep network. Interpretation, alignment, and disentanglement of latent spaces are important in different computer vision techniques such as unsupervised learning [30], multi-modal information fusion [24, 31], generative models [32, 33], knowledge distillation [34, 35], and transfer learning [36]. However, the interpretation of latent space is challenging because it usually requires implicit domain knowledge based on human judgment [37].

To interpret the latent space, researchers either apply dimension reduction tools such as PCA [38], T-SNE [13], or develop interactive analysis tools [39]. To manipulate latent space, a variation autoencoder (VAE) can be used to map the input to latent space and disentangle/match the different latent spaces [40, 41]. In this work, we utilize the intrinsic property of the connectivity mask and propose a simple yet effective sub-path slicing-based method to disentangle the directional subspace from the shared latent space, followed by a visual demonstration of the effectiveness of our disentanglement process with T-SNE.

### 2.3. Self-attention mechanism

Self-attention is widely used in computer vision as it can capture the dependencies between latent features. The self-attention mechanism is defined as:

$$y = f(\alpha, g(\text{x})), \qquad (1)$$

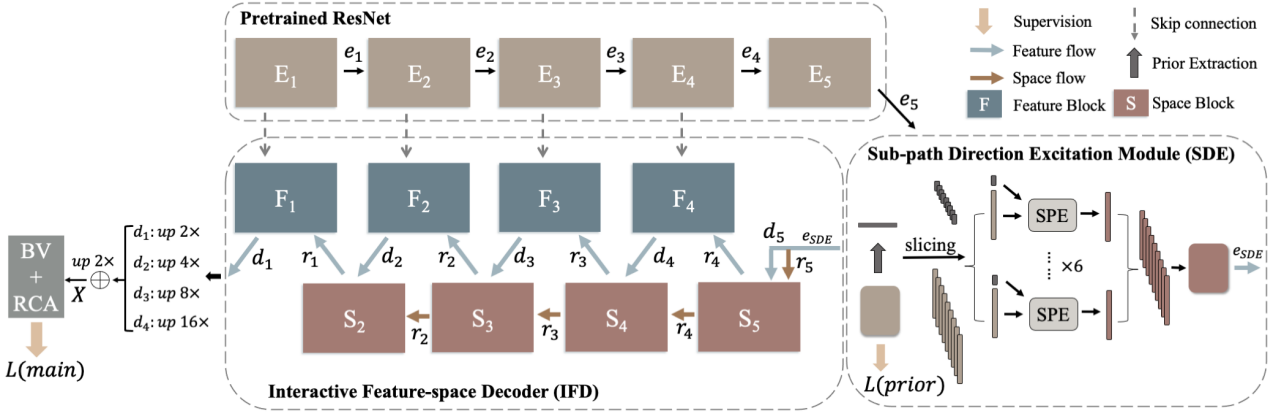

Figure 4. The overview of the proposed DconnNet. It contains three parts: a pretrained ResNet encoder, a Sub-path Direction Excitation Module, and an Interactive Feature-space Decoder. The term "$up\ N\times$" means upsampling with a stride of $N$.

where $x$ is the input feature map, $g$ is a unary function that computes the embedding of input, and $f$ is usually a matrix multiplication [10, 42] or a simple dot product [11, 43, 44], depending on the definition of $\alpha$. $\alpha$ is the attention/relation map which can be a mutual relation map by spatial- or channel-wisely computing the similarity between one pixel with other pixels in a feature map, or between two related feature maps from the same feature space [10, 42]; it can be a contextual-spatial relation map generated by similarity estimation between a feature map and a latent vector with abstract contextual meaning [44, 45]. Broadly speaking, $\alpha$ can also be a single map [11] or a vector [43, 46] containing special contextual meaning.

## 3. Method

### 3.1. Directional space in connectivity modeling

Due to connectivity between different pixel classes and directions, there are two groups of features in the latent space of a connectivity-based network: categorical, directional. Each group of features forms its specific subspace in the latent space. In a single-path connectivity network, these two subspaces are highly coupled (Fig. 2), resulting in low-discriminative features. We demonstrate that the efficient disentangling and effective utilization of the direction space can enhance the overall feature representation in the connectivity model.

In a connectivity mask, different channels represent different directions of the pixel connection. Thus, as the network goes deeper, it naturally stores directional information among channels. Based on this property, the directional features can be captured and manipulated through channel-wise operations. Specifically, we propose SDE to disentangle channel-wise directional features from the latent space, followed by IFD to extract the directional embeddings at different layers and use them to enhance the overall feature representation in a self-attention manner. The overall DconnNet structure is shown in Fig. 4.

### 3.2. Sub-path Direction Excitation Module

**Directional prior extraction**. Taking advantage of the channel-wise directional information in the connectivity mask, a direct way to get distinctive directional embedding is to coarsely supervise the intermediate features and squeeze the channels of the auxiliary connectivity output. We denote the encoder's outputs as $e_n$, where $n$ is the $n_{th}$ encoder layer. As shown in Fig. 5, we upsample $e_5$, the last encoder output, to the input size and get a preliminary output called $X_{prior}$, which will be supervised to learn the connectivity mask in the loss calculation. Therefore, rich, distinctive directional information can be obtained from the channels of $X_{prior}$. Then, we squeeze [43] $X_{prior}$ through a global average pooling (GAP) and map the vector to the same dimension as the latent feature map $e_5 \in R^{C_{e_5} \times H_{e_5} \times W_{e_5}}$ with a 1×1 convolutional kernel $W_1$ [44]:

$$GAP(X) = \frac{1}{H \times W} \sum_{i=1}^{H} \sum_{j=1}^{W} X(i,j), \quad (2)$$

$$v_{prior} = \delta\left(W_1 GAP(X_{prior})\right), \quad (3)$$

where H and W are the height and the width of a feature map, $W_1 \in R^{C_{e_5} \times C_k}$, $C_k$ is the channel of $X_{prior}$, and $\delta$ is the ReLu activation. The resulting $v_{prior} \in R^{C_{e_5}}$ has the directional information of a specific direction embedded in each entry. Next, we re-encode $v_{prior}$ with a 1×1 convolution $W_2 \in R^{C_{e_5} \times C_{e_5}}$ and apply a sigmoid gating function $\sigma$ to normalize the projected vector:

$$\alpha_{prior} = \sigma(W_2 v_{prior}). \quad (4)$$

Since $\alpha_{prior}$ contains rich element-wise directional information, we call it the directional prior.

**Channel-wise slicing**. To early disentangle the categorical and directional subspaces in the hidden layers, we split the latent features ($e_5$) and the directional prior ($\alpha_{prior}$) into eight parts by channel-wise slicing. We denote the $i_{th}$ slices as $e_5^i$ and $\alpha_{prior}^i$.

**Sub-path excitation (SPE)**. We construct a sub-path for each pair of these feature-embedding slices. In each sub-path, we pass the feature slice $e_5^i$ through spatial and channel attention modules [10] to capture the long-range and inter-channel dependencies, resulting in $e_5^{i\prime}$. Next, we channel-wisely multiply the $\alpha_{prior}^i$ with $e_5^{i\prime}$ to selectivity highlight or suppress features with specific directional information. Then, we recode the output with a 1×1 convolution kernel $W_3^i$ and residually output it as:

$$e_{SDE}^i = W_3^i(\alpha_{prior}^i \cdot e_5^{i\prime}) + e_5^i. \quad (5)$$

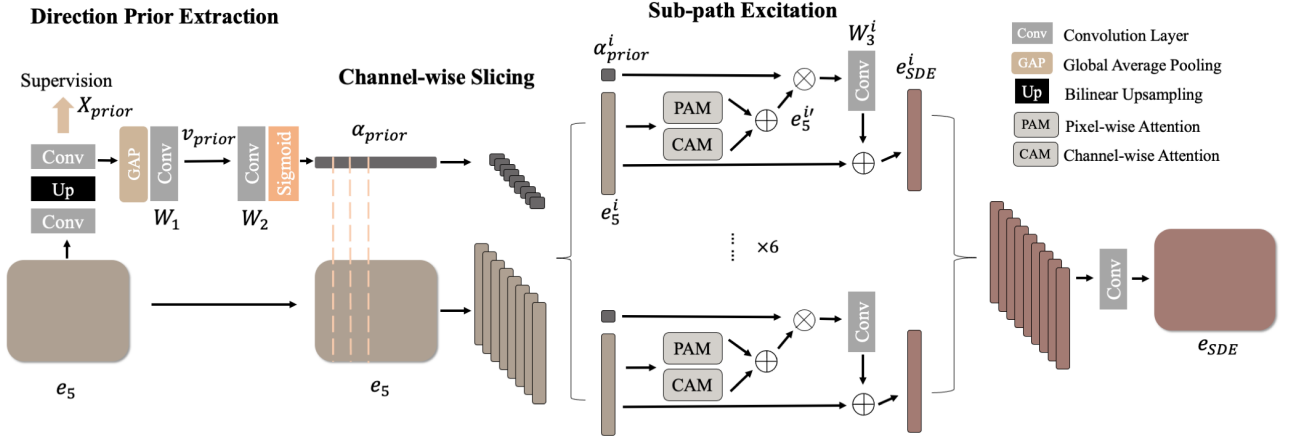

Figure 5. The SDE module, including three steps: direction prior extraction, channel-wise slicing, and sub-path excitation.

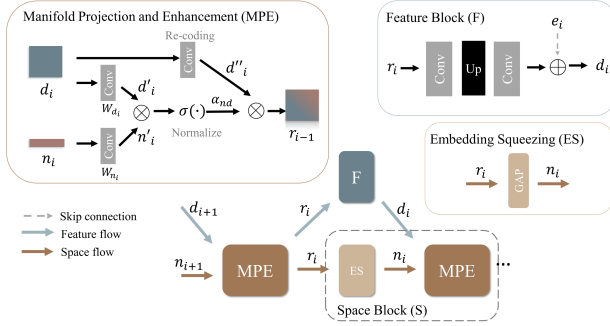

Figure 6. The Interactive Feature-space Decoder (IFD).

Finally, we stack all sub-paths outputs ($e_{SDE}^i$) into one and recode it, resulting in a new feature map, $e_{SDE}$.

Due to the slicing operation, each group of slices will only contain part of the full features. However, the shrink of discriminative contextual information in directional and categorical features differs. Specifically, $\alpha_{prior}$ is a highly discriminative directional embedding as it is a low-level linear combination of unique directional features. Thus, channel-wise slicing will cause significant shrinkage of directional information in each slice. However, $e_5$ contains a group of high-level but less-discriminative categorical features which usually have low variations with others [47]. As a result, high channel-wise categorical correlation and redundancy [48] exist in $e_5$. Therefore, a channel-wise slicing will result in less shrinkage [49] of high-discriminative categorical features in $e_5^i$. By doing this, we contextual-unevenly divide the directional and categorical features in each sub-path and make each sub-path shift its focus to the dominant features. Specifically, inside each sub-path, the network learns how to focus on the dominant class-specific information with less distinctive directional information emphasized in the channels. Between sub-paths, the network learns different distinctive directional information. Then, once we stack the sub-paths back, the direction information will be naturally disentangled from the original latent space and embedded into the channels.

### 3.3. Interactive Feature-space Decoder

To ensure that the direction-dependent information can be effectively fused into feature maps in each layer, we propose an Interactive Feature-space Decoder with two top-down dynamically interactive flows, i.e., space flow and feature flow, as in Fig. 6.

**Feature maps**. There are two types of feature maps in each layer of IFD: the main feature map $d_i$ and the direction-enhanced map $r_i$. At the first decoder layer ($i = 5$), they are both initialized as the outputs of SDE:

$$r_5 = d_5 = e_{SDE}, \quad (6)$$

and are updated separately in the space and feature flows.

**Space flow**. In each space flow, a Space Block (S), which contains the embedding squeezing (ES) module and the manifold projection and enhancement (MPE) module, is used to enhance the directional representation. The ES module takes $r_i$ as the input and outputs a directional embedding $n_i$, a high-level directional representation as:

$$n_i = GAP(r_i). \quad (7)$$

Then, we use $n_i$ as the input of MPE to enhance the directional representation in the main feature map $d_i$. In MPE, we first project both the main feature map $d_i \in R^{C_{d_i} \times H_{d_i} \times W_{d_i}}$ and directional embedding $n_i \in R^{C_{d_i}/2}$ onto a shared manifold $R^{C_{d_i}}$ at the current resolution with two 1×1 convolutional projectors:

$$d_i' = W_{d_i} d_i, n_i' = W_{n_i} n_i, \quad (8)$$

where $W_{d_i} \in R^{C_{d_i} \times C_{d_i}}$ and $W_{n_i} \in R^{C_{d_i} \times C_{d_i}/2}$. Next, we model the category-direction relation by calculating the similarity between the projected directional embedding $n'_i$ and feature map $d'_i$ with a channel-wise dot product and a sigmoid activation:

$$\alpha_{nd} = \sigma(d'_i \cdot n'_i), \quad (9)$$

where $\alpha_{nd}$ is the normalized category-direction attention map in which the direction-relevant features are enhanced and the irrelevant features are suppressed across channels. Next, we enhance the directional information in the re-encoded original feature map $d''_i$ with the attention map $\alpha_{nd}$ using a point-wise inner multiplication:

$$r_{i-1} = \alpha_{nd} \cdot d''_i. \quad (10)$$

By doing this, we effectively fuse the directional information into the feature map, resulting in a new direction-enhanced map $r_{i-1}$, which will be further used as the input of feature flow and to generate the directional embedding $n_{i-1}$ for the next layer.

**Feature flow**. Each feature flow contains a Feature Block (F), including two convolutional layers and an upsampling layer with a skip connection. It takes $r_i$ as the input and outputs the upsampled main feature map $d_i$.

This top-down design and dynamic interactions between the two flows are mutually beneficial. Specifically, the embedding $n_i$ extracted by space flow effectively fuses the directional information into the feature map $d_i$ from the feature flow. On the other hand, the $d_i$ generated by feature flow refines the directional representation $n_{i-1}$ by adding supplementary directional information at a higher resolution to the space flow via MPE and ES.

### 3.4. Connectivity output

We integrate the main feature maps $d_1$ to $d_4$ to get the final connectivity output $X$. Every eight channels of $X$ represent the connectivity of one class. In line with [16], we use the Bilateral Voting (BV) module in (11) and the Region-guided Channel Aggregation (RCA) module in (12) to get the final segmentation map.

$$\tilde{X}_j(x,y) = \tilde{X}_{9-j}(x+a, y+b)$$
$$= X_j(x,y) \times X_{9-j}(x+a, y+b), \quad (11)$$
$$\tilde{S}(x,y) = max\{\tilde{X}_i(x,y)\}_{i=1}^{8}, \quad (12)$$

where j is the $j_{th}$ channel, $a, b \in \{0, \pm 1\}$ represent the location of neighboring pixel, $\tilde{X}$ is the Bicon map, $\tilde{S}$ is the final segmentation prediction [16].

### 3.5. Loss function

The total loss function of our work is:

$$L_{total} = L(main) + 0.3 * L(prior), \quad (13)$$

where $main$ represents the main output and $prior$ represents the auxiliary output from SDE. Each loss $L$ contains two parts, the Size Density Loss $L_{sd}$ and the original Bicon loss [16] $L_{bicon}$:

$$L = L_{sd} + L_{bicon}. \quad (14)$$

**Size Density Loss.** Considering the imbalanced nature of the medical data, we propose a novel imbalanced loss function based on the label size distribution of the classes of interests in the dataset, called Size Density Loss.

Before training, we sample all the training images and calculate the size of the label (total positive pixel number) on each image. Based on the extracted size distribution, for each class $j$, we calculate the probability density function of label size $k$, i.e., $PDF_j(k)$. Then, in the training phase, for each training image, we calculate its label size for each class $j$ and get its size density weight $P_j(k)$ as:

$$P_j(k) = \begin{cases} 1, & k = 0, \\ -log\left(PDF_j(k)\right), & k \neq 0. \end{cases} \quad (15)$$

We define the $L_{sd}$ as a variant of Dice loss [50]:

$$L_{sd} = \sum_j^{Class} P_j(k) \left(1 - \frac{2 \times \sum(S \times G_s) + \varepsilon}{\sum S + \sum G_s + \varepsilon}\right), \quad (16)$$

where $S$ is the final segmentation prediction, $\varepsilon$ is the stabilization term [51] and is usually set as 1.

**Bicon Loss term.** We use the Bicon Loss $L_{bicon}$ for connectivity modeling, as originally defined in [16]:

$$L_{bicon} = L_{decouple} + L_{con\_const}. \quad (17)$$

## 4. Experiments

### 4.1. Datasets and evaluation metrics

We used three popular and diverse medical benchmark datasets in this paper. Specifically, we used Retouch [52] and ISIC2018 [53] datasets to evaluate the DconnNet performance for large-scale medical segmentation. We used CHASEDB1 [54] dataset to assess the topological performance of the DconnNet.

**Retouch** is an OCT retinal fluids segmentation benchmark that has been widely used [55, 56] for assessing computer vision methods. It contains three classes of disease biomarkers: intraretinal fluid (IRF), subretinal fluid (SRF), and pigment epithelial detachment (PED). Retouch contains a training set of 70 OCT volumes from three OCT scanners, with frame sizes spanning from 512 ×496 to 512 × 1024 pixels. Retouch is a two-level balanced dataset: at the image level, biomarkers do not span the whole volume; at the pixel level, each biomarker has a relatively small size compared to the background. Since the testing labels are unavailable, we implemented the volume-level 3-fold cross-validation (CV) for each scanner on the training data set. Following the official guidelines [52], we used volume-level Dice [65] ($DSC_v$), image-level Dice ($DSC$), absolute volume difference ($AVD$), and volume-wise balanced accuracy [66] ($BACC$) to evaluate the results.

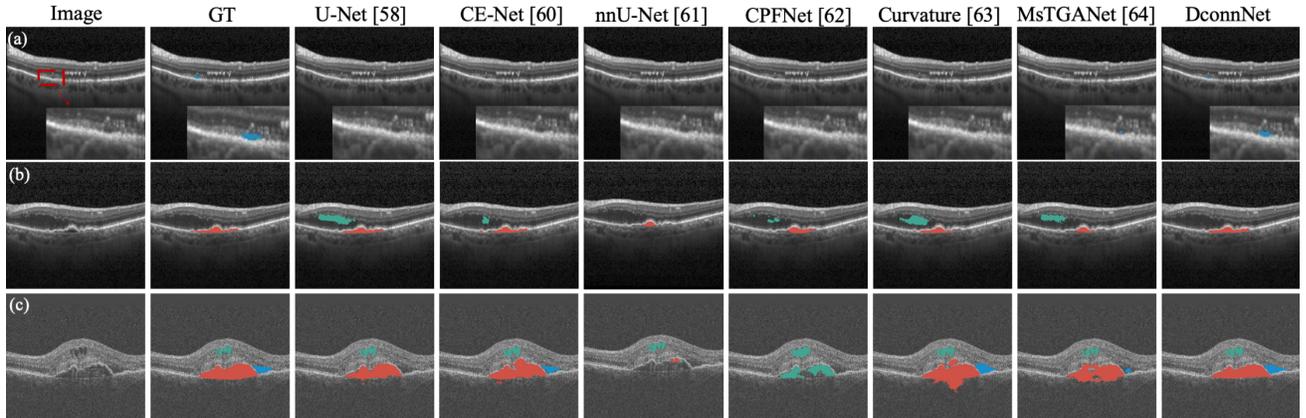

Figure 7. Visual comparison between the proposed DconnNet and other state-of-the-art methods on Retouch dataset. Different colors of masks represent different biomarker classes. Green: IRF; blue: SRF; red: PED.

Table 1. Results on Retouch dataset. Model size (M) and testing speed (FPS) are also reported. The best results are **bold**.

| Method | $DSC_v$ | DSC | AVD | BACC | Size / Speed |
|---|---|---|---|---|---|
| DeepLabv3+ [57] | 60.2 | 82.7 | 0.023 | 86.5 | 59.3 / 38 |
| U-Net [58] | 66.1 | 84.2 | 0.021 | 87.4 | 13.4 / 38 |
| Att-UNet [59] | 65.3 | 83.4 | 0.022 | 86.6 | 34.9 / 36 |
| CE-Net [60] | 67.3 | 84.2 | 0.026 | 84.6 | 29.0 / 37 |
| nnU-Net [61] | 67.2 | 84.3 | 0.023 | 86.4 | 30.0 / 20 |
| CPFNet [62] | 69.0 | 85.7 | 0.022 | 88.0 | 43.3 / 37 |
| Curvature [63] | 68.2 | 84.7 | 0.024 | 87.1 | 43.3 / 37 |
| MsTGANet [64] | 68.9 | 85.0 | 0.023 | 87.1 | 11.6 / 37 |
| DconnNet | **78.2** | **87.7** | **0.020** | **90.5** | 36.4 / 40 |

**ISIC 2018** is a popular medical segmentation benchmark [67-70] containing 2594 images of various types of skin lesions at different resolutions. Following [71], we used 5-fold CV and evaluated the results with Dice, IOU, accuracy (ACC), and precision (PREC) metrics.

**CHASEDB1** is a vessel segmentation dataset containing 28 fundus images with a resolution of 999× 960 pixels and two sets of manual annotations. The first manual annotation is adopted in this work as in [72]. We conducted 5-fold CV and evaluated the results with two volumetric metrics *Dice* and *IOU* ; and three topology-similarity-based metric *clDice*, 0- and 1-Betti numbers ($\beta_0$ and $\beta_1$) [73] to measure the topological similarity.

### 4.2. Experimental details

We generated the connectivity mask for each image via simple matrix operations [16]. We used the ImageNet [74] pretrained ResNet [75] as our encoder. The hyperparameter setting differed across datasets. For the Retouch dataset, we resized each image to 256×256 with no data augmentation and trained DconnNet with a 'poly' learning rate strategy. For ISIC2018, we resized images to 224×320 and used the same data augmentation and the parameter settings in [71]. For CHASEDB1, to fully use the limited data, we resized to 960×960 pixels and applied the same augmentations as in [60]. Since the CHASEDB1 dataset has a limited sample size, we did not apply SDL loss to this dataset. We did not use any pre-processing for training and no post-processing for evaluation. The framework is built on PyTorch 1.7.0 [76]. All experiments are performed with a GPU NVIDIA GeForce GTX 3090 Ti.

## 5. Results

### 5.1. Retinal fluid segmentation

**Comparison of state-of-the-art methods**. We compared the proposed DconnNet with eight state-of-the-art models for relevant applications, including DeepLab V3+ [57], U-Net [58], Attention UNet [59], CE-Net [60], nnU-Net [61], CPFNet [62], CPFNet-backboned Curvature Loss [63], and MsTGANet [64] in Table 1, which shows the superior performance of the proposed model. Specifically, the improvements on $DSC_i$ and $DSC_v$ demonstrate that our model has a consistent prediction on the small fluid regions, while the improvements on AVD and BACC demonstrate that our model has an overall more accurate prediction on the fluid regions, regardless of the target size.

**Qualitative study**. In Fig. 7, we compared DconnNet with the top-six baselines. In image (a), all competing methods missed the tiny SRF region, while our DconnNet made an accurate prediction. In (b), all methods, except DconnNet, either made false positive (FP) segmentations of IRF or an incomplete segmentation on PED. In (c), only our DconnNet made the topologically connected prediction for PED and accurately segmented the IRF and SRF regions.

### 5.2. Skin lesion segmentation

**Comparison of state-of-the-art methods**. In Table 2, we compared our method with U-Net [58], BCDU-Net [77], CE-Net [60], nnU-Net [61], HiFormer [78], CPFNet [62], FATNet [79], and Ms RED [71]. The proposed DconnNet outperformed all competitive models.

**Qualitative study**. Due to limited space, in Figs. 8 and 9,

Table 2. Results on ISIC2018 dataset. The best results are **bold**.

| Method | DSC | IOU | ACC | PREC |
|---|---|---|---|---|
| U-Net [58] | 88.41 | 81.23 | 95.53 | 90.7 |
| BCDU-Net [77] | 88.33 | 80.84 | 95.48 | 89.68 |
| CE-Net [60] | 89.23 | 82.34 | 95.76 | 91.51 |
| nnU-Net [61] | 89.24 | 82.35 | 95.79 | 91.45 |
| HiFormer [78] | 88.54 | 81.45 | 95.59 | 91.09 |
| CPFNet [62] | 89.34 | 82.64 | 95.89 | 91.38 |
| FATNet [79] | 88.84 | 81.79 | 95.62 | 91.18 |
| Ms RED [71] | 89.48 | 82.71 | 95.89 | **91.83** |
| DconnNet | **90.43** | **83.91** | **96.39** | 91.54 |

Table 3. Results on CHASEDB1. The best results are **bold**.

| Method | clDice | DSC | IOU | $\beta_0$ | $\beta_1$ |
|---|---|---|---|---|---|
| U-Net [58] | 74.0 | 74.7 | 59.3 | 1.390 | 2.633 |
| Att-UNet [59] | 75.3 | 75.7 | 61.0 | 1.330 | 2.531 |
| GT-DLA [80] | 81.0 | 80.6 | 67.4 | 0.790 | 1.969 |
| CE-Net [60] | 82.0 | 81.0 | 68.1 | 0.383 | 1.670 |
| clDiceLoss [73] | 82.9 | 81.0 | 68.1 | 0.345 | 1.656 |
| GraphCutLoss [72] | 82.6 | 81.4 | 68.8 | 0.437 | 1.692 |
| DconnNet | **83.3** | **81.8** | **69.4** | **0.341** | **1.630** |

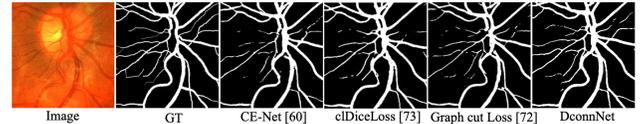

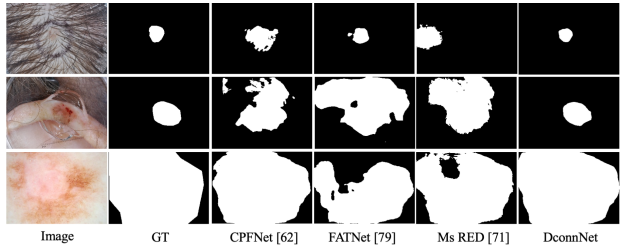

Figure 8. Visualization on ISIC2018 dataset.

Figure 9. Visualization on CHASEDB1 dataset.

Table 4. Ablation study of the Retouch Dataset. Conn stands for connectivity modeling with $L_{bicon}$. DS stands for dice loss.

| Conn | Module | | Loss | | DSC | $DSC_v$ | AVD | BACC |
| | SDE | IFD | DS | SDL | | | | |
|---|---|---|---|---|---|---|---|---|
| 1 | | | √ | | 83.2 | 65.0 | 0.025 | 87.0 |
| 2 | √ | | √ | | 85.8 | 70.3 | 0.024 | 87.7 |
| 3 | √ | √ | √ | | 86.2 | 73.7 | 0.023 | 87.9 |
| 4 | √ | √ | √ | | 86.7 | 75.2 | 0.023 | 88.3 |
| 5 | √ | √ | √ | √ | 87.7 | 78.2 | 0.020 | 90.5 |

we only compare the best-performing methods. In Fig. 8, DconnNet can accurately segment the lesion even under strong clinical confounders (e.g., the hairs in the first image). Also, DconnNet could learn the topology of skin lesions and make connected predictions in all cases, while others showed topological errors.

### 5.3. Topological vessel segmentation

We used this experiment to demonstrate the topology-preserving ability of DconnNet in vessel segmentation, in which anatomical consistency is crucial. Due to the limited sampling size, we did not use SDL in this experiment.

**Comparison of state-of-the-art methods**. We compared the results of DconnNet versus up-to-date leading methods, including U-Net [58], Att-UNet [59], GT-DLA [80], CE-Net [60], clDiceLoss [73], and Graph Cut Loss [72] in Table 3. Our method surpassed other methods on all metrics. The superior performances of DconnNet on *clDice*, $\beta_0$, and $\beta_1$, reflect the topological similarity between predictions and labels.

**Qualitative study on the vessel topology**. In Fig. 9, we visually compared different methods. While in part due to limited training data DconnNet could not make perfect predictions across the whole images, it visually outperformed other methods. Our model's predicted vessels are mostly connected and with no significant topological errors. Other models showed more topological issues, e.g., in the form of non-simply connected regions.

### 5.4. Ablation study

We conducted ablation studies to compare directional connectivity modeling with the traditional segmentation-based modeling and the naïve connectivity modeling [16]. All experiments in this section are on the Retouch Dataset.

**Overall ablation study**. The overall ablation study is reported in Table 4. The backbone (Exp. 1), a pretrained ResNet [75] encoder with a regular decoder, is from [62]. Exp. 2 is identical to [16] with a different backbone. Both connectivity modeling (Exp. 2) and directional modeling (Exp. 3 and 4) result in significant performance improvements. Especially, the considerable increases in the $DSC$ terms demonstrate that the model is becoming stable when dealing with the small fluid regions. Finally, by adding SDL (Exp. 5), all metrics got improved showing the impact of the distribution-based weighting scheme.

**Directional prior in SDE**. The directional prior in the SDE module was designed to provide an initial prior to disentangle the directional features which will later be utilized across the entire network. Thus, Table 5 compares two experiments: the complete DconnNet and the one without directional prior, i.e., only PAM and CAM in each sub-path of SDE. Even when most of the network structure was kept unchanged, the network performed worse without the guidance of an initial directional embedding prior.

**Sub-path attention vs. single-path attention**. In SDE, we applied a sub-path attention unit in each slice to capture the dependency between the contextual-unevenly sliced directional and categorical features. To demonstrate the effectiveness of the sub-path attention, we compared our proposed SDE with the alternative single-path attention units. We first compared it with a regular dual attention unit [10] which is the direct alternative to our SDE since we used the same module in each sub-path. We also conducted an experiment with the non-local unit [42]. The results are shown in Table 6, which shows all metrics increase after introducing the sub-path mechanism.

Table 5. Ablation study on directional prior.

| DconnNet | DSC | $DSC_v$ | AVD | BACC |
|---|---|---|---|---|
| w/o. prior | 86.3 | 72.3 | 0.023 | 89.1 |
| w/. prior | **87.7** | **76.6** | **0.020** | **90.5** |

Table 6. Ablation study on Sub-path attention. Backbone_Conn is the connectivity-based modeling on backbone [16], DA is the dual attention, and NL is the non-local module.

| Backbone_Conn | DSC | $DSC_v$ | AVD | BACC |
|---|---|---|---|---|
| + DA [10] | 84.7 | 69.5 | 0.027 | 86.2 |
| + NL [42] | 85.1 | 70.0 | 0.028 | 87.2 |
| + SDE | **87.7** | **76.6** | **0.020** | **90.5** |

**Disentanglement of directional subspace.** To show that the directional subspace is disentangled from the shared latent space with the proposed sub-path slicing method, we conducted two experiments on the DconnNet's latent space. First, we used T-SNE to visualize the learned patch-wise feature maps in the latent spaces before and after the SDE module of a trained DconnNet, as shown in Fig. 2. Applying SDE decoupled the latent space and changed the latent features' centered distribution to a polarized distribution. We interpret this as the decoupling process between the directional subspace and categorical subspace.

Then, to show the well-structured directional subspace between channels, we visualize the learned channel embeddings in the latent spaces before and after the SDE module of a trained DconnNet, as in Fig. 10. After SDE, the channel embeddings naturally grouped into several distinctive parts, demonstrating the effectiveness of the sub-path excitation.

**Comparison of size density loss.** Exp. 4 and 5 in Table 4 demonstrated the superiority of the proposed SDL in connectivity modeling. This subsection further analyzes SDL in a pixel-classification-based segmentation setting.

In Table 7, we compared SDL with alternative loss functions with a similar idea of weighting the region-based Dice loss, including Exponential Logarithmical Loss [81], Focal Dice Loss [82], Dice loss [50], and Generalised Dice [83] based on two networks: U-Net and CPFNet. Our SDL achieves the highest $DSC$ both image- and volume-wise. Moreover, it shows more stability when dealing with a two-level imbalanced medical dataset compared to feedback-based losses (e.g., Focal-like losses) since it is less sensitive to FP prediction on the negative training images.

**Model size.** Table 1 compares different model sizes. DconnNet (36.42M parameters) performed far better than the backbone (33.16M) with only ~3M extra parameters. In SDE, due to the channel slicing, we quadratically reduced the size of each sub-path attention (0.196M) from the full-size dual attention [10] (23.35M), resulting in the small size of the proposed SDE in Table 8. Also, changing the network to connectivity modeling only takes extra 0.03M parameters. Given the relatively small parameter increase, 3D connectivity modeling is a potential future direction for 3D image segmentation.

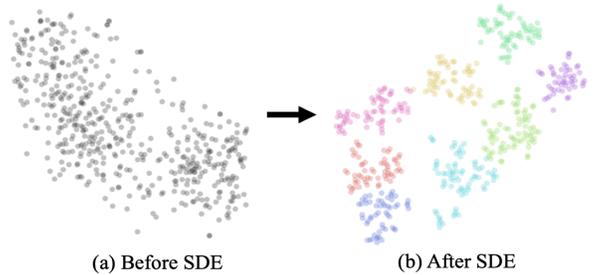

(a) Before SDE      (b) After SDE

Figure 10. Visualization of latent channel embeddings of DconnNet before and after SDE module using T-SNE. The colors in (b) indicate the unsupervised clustering result. When applied to SDE, the channel embeddings naturally grouped into several distinctive parts.

Table 7. Ablation study on loss function on Retouch Dataset.

| Loss | U-Net | | CPFNet | |
|---|---|---|---|---|
| | DSC | $DSC_v$ | DSC | $DSC_v$ |
| Exponential [81] | Not Converged | | | |
| Focal [82] | | | | |
| Dice [50] | 82.9 | 65.6 | 83.1 | 65.2 |
| Generalised [83] | 80.1 | 54.1 | 80.9 | 55.5 |
| SDL | **85.1** | **69.2** | **85.4** | **69.1** |

Table 8. The components' size of DconnNet.

| Module | ResNet Encoder | SDE | IFD Feature Block | IFD Space Block | Final Decoder | Connect. Modeling | Total |
|---|---|---|---|---|---|---|---|
| Para. (M) | 21.80 | 1.57 | 11.35 | 1.49 | 0.18 | 0.03 | 36.42 |

## 6. Conclusion

This paper proposed a novel directional connectivity modeling network (DconnNet) for medical image segmentation. The core idea is to disentangle the directional subspace from the shared latent space and use the extracted directional features to enhance the overall data representation. We demonstrated the effectiveness of DconnNet in three ways. First, by statistical comparisons to other state-of-the-art methods, we showed the overall better performance of DconnNet. Then, we demonstrated its topology-preserving ability by qualitatively and quantitively comparing DconnNet to other methods on a topologically sensitive dataset to other methods. Third, by visualizing the latent space of DconnNet, we revealed the disentanglement process of the directional subspace.

## Acknowledgment

This research was supported in part by NIH R01 EY030124, U01 EY034687, R01 EY031033, UG1 EY033287, and R01 AG072732.

# References

[1] P. L. Bazin, and D. L. Pham, "Topology correction of segmented medical images using a fast-marching algorithm," *Comput. Methods Programs Biomed.*, vol. 88, no. 2, pp. 182-90, 2007.

[2] J. R. Clough, N. Byrne, I. Oksuz, V. A. Zimmer, J. A. Schnabel, and A. P. King, "A topological loss function for deep-learning based image segmentation using persistent homology," *IEEE TPAMI*, vol. 44, no. 12, pp. 8766-8778, 2022.

[3] X. Hu, F. Li, D. Samaras, and C. Chen, "Topology-preserving deep image segmentation," in *NIPS*, vol. 32, 2019.

[4] T. Y. Kong, and A. Rosenfeld, "Digital topology: Introduction and survey," *Comput. Gr. Image Process.*, vol. 48, no. 3, pp. 357-393, 1989.

[5] P. K. Saha, R. Strand, and G. Borgefors, "Digital topology and geometry in medical imaging: a survey," *IEEE Trans. Med. Imag.*, vol. 34, no. 9, pp. 1940-1964, 2015.

[6] S. K. Nath, K. Palaniappan, and F. Bunyak, "Cell segmentation using coupled level sets and graph-vertex coloring," in *MICCAI*, pp. 101-108, 2006.

[7] N. Kriegeskorte, and R. Goebel, "An efficient algorithm for topologically correct segmentation of the cortical sheet in anatomical MR volumes," *Neuroimage*, vol. 14, no. 2, pp. 329-346, 2001.

[8] M. Liu, F. Colas, and R. Siegwart, "Regional topological segmentation based on mutual information graphs," in *ICRA*, pp. 3269-3274, 2011.

[9] J. Long, E. Shelhamer, and T. Darrell, "Fully convolutional networks for semantic segmentation," in *CVPR*, pp. 3431-3440, 2015.

[10] J. Fu, J. Liu, H. Tian, Y. Li, Y. Bao, Z. Fang, and H. Lu, "Dual Attention Network for Scene Segmentation," in *CVPR*, pp. 3141-3149, 2019.

[11] H. Zhao, Y. Zhang, S. Liu, J. Shi, C. C. Loy, D. Lin, and J. Jia, "PSANet: Point-wise Spatial Attention Network for Scene Parsing," in *ECCV*, pp. 270-286, 2018.

[12] L. C. Chen, G. Papandreou, I. Kokkinos, K. Murphy, and A. L. Yuille, "DeepLab: Semantic Image Segmentation with Deep Convolutional Nets, Atrous Convolution, and Fully Connected CRFs," *IEEE TPAMI*, vol. 40, no. 4, pp. 834-848, 2018.

[13] L. Van der Maaten, and G. Hinton, "Visualizing data using t-SNE," *JMLR*, vol. 9, no. 11, 2008.

[14] H. Liu, F. Liu, X. Fan, and D. Huang, "Polarized self-attention: Towards high-quality pixel-wise mapping," *Neurocomputing*, vol. 506, pp. 158-167, 2022.

[15] M. Kampffmeyer, N. Dong, X. Liang, Y. Zhang, and E. P. Xing, "ConnNet: A long-range relation-aware pixel-connectivity network for salient segmentation," *IEEE Trans. Image Process.*, vol. 28, no. 5, pp. 2518-2529, 2018.

[16] Z. Yang, S. Soltanian-Zadeh, and S. Farsiu, "BiconNet: An Edge-preserved Connectivity-based Approach for Salient Object Detection," *Pattern Recognit.*, vol. 121, pp. 108231, 2022.

[17] L. Borne, D. Rivière, M. Mancip, and J.-F. Mangin, "Automatic labeling of cortical sulci using patch- or CNN-based segmentation techniques combined with bottom-up geometric constraints," *Med. Image Anal.*, vol. 62, pp. 101651, 2020.

[18] A. Rosenfeld, and R. Klette, "Digital geometry," *Information Sciences*, vol. 148, no. 1, pp. 123-127, 2002.

[19] X. Li, Y. Wang, L. Zhang, S. Liu, J. Mei, and Y. Li, "Topology-Enhanced Urban Road Extraction via a Geographic Feature-Enhanced Network," *IEEE Trans. Geosci. Remote Sens.*, vol. 58, no. 12, pp. 8819-8830, 2020.

[20] H. Ma, C. Li, J. Liu, J. Wang, and M. Q. H. Meng, "Enhance Connectivity of Promising Regions for Sampling-Based Path Planning," *IEEE Trans. on Autom. Sci. Eng.*, early access. doi: 10.1109/TASE.2022.3191519, 2022.

[21] Y. Qin, M. Chen, H. Zheng, Y. Gu, M. Shen, J. Yang, X. Huang, Y.-M. Zhu, and G.-Z. Yang, "Airwaynet: a voxel-connectivity aware approach for accurate airway segmentation using convolutional neural networks," in *MICCAI*, pp. 212-220, 2019.

[22] Z. Yang, S. Soltanian-Zadeh, K. K. Chu, H. Zhang, L. Moussa, A. E. Watts, N. J. Shaheen, A. Wax, and S. Farsiu, "Connectivity-based Deep Learning Approach for Segmentation of the Epithelium in In Vivo Human Esophageal OCT Images," *Biomed. Opt. Express*, vol. 12, no. 10, pp. 6326-6340, 2021.

[23] Y. Jia, M. Salzmann, and T. Darrell, "Factorized latent spaces with structured sparsity," in *NIPS*, vol. 23, 2010.

[24] J. Gu, Z. Wang, W. Ouyang, J. Li, and L. Zhuo, "3d hand pose estimation with disentangled cross-modal latent space," in *WACV*, pp. 391-400, 2020.

[25] M. Salzmann, C. H. Ek, R. Urtasun, and T. Darrell, "Factorized orthogonal latent spaces," in *AISTATS*, pp. 701-708, 2010.

[26] A. K. Rudra, A. S. Chowdhury, A. Elnakib, F. Khalifa, A. Soliman, G. Beache, and A. El-Baz, "Kidney segmentation using graph cuts and pixel connectivity," *Pattern Recognit. Lett.*, vol. 34, no. 13, pp. 1470-1475, 2013.

[27] M. Fontaine, L. Macaire, and J.-G. Postaire, "Image segmentation based on an original multiscale analysis of the pixel connectivity properties," in *ICIP*, pp. 804-807, 2000.

[28] Y. Long, D. Morris, X. Liu, M. Castro, P. Chakravarty, and P. Narayanan, "Radar-camera pixel depth association for depth completion," in *CVPR*, pp. 12507-12516, 2021.

[29] R. Rasti, A. Biglari, M. Rezapourian, Z. Yang, and S. Farsiu, "RetiFluidNet: A Self-Adaptive and Multi-Attention Deep Convolutional Network for Retinal OCT Fluid Segmentation," *IEEE Trans. Med. Imag.*, (In Press), 2023.

[30] Y. Cai, K.-Y. Lin, C. Zhang, Q. Wang, X. Wang, and H. Li, "Learning a Structured Latent Space for Unsupervised Point Cloud Completion," in *CVPR*, pp. 5543-5553, 2022.

[31] L. Yang, and A. Yao, "Disentangling latent hands for image synthesis and pose estimation," in *CVPR*, pp. 9877-9886, 2019.

[32] G. Arvanitidis, L. K. Hansen, and S. Hauberg, "Latent space oddity: on the curvature of deep generative models," in ICLR, 2018.

[33] Z. Li, R. Tao, J. Wang, F. Li, H. Niu, M. Yue, and B. Li, "Interpreting the latent space of GANs via measuring decoupling," *IEEE Trans. Artif. Intell.*, vol. 2, no. 1, pp. 58-70, 2021.

[34] C. Yang, H. Zhou, Z. An, X. Jiang, Y. Xu, and Q. Zhang, "Cross-image relational knowledge distillation for semantic segmentation," in *CVPR*, pp. 12319-12328, 2022.

[35] M. Hu, M. Maillard, Y. Zhang, T. Ciceri, G. La Barbera, I. Bloch, and P. Gori, "Knowledge distillation from multi-

modal to mono-modal segmentation networks," in *MICCAI*, pp. 772-781, 2020.
[36] B. Delhaisse, D. Esteban, L. Rozo, and D. Caldwell, "Transfer learning of shared latent spaces between robots with similar kinematic structure," in *IJCNN*, pp. 4142-4149, 2017.
[37] Y. Liu, E. Jun, Q. Li, and J. Heer, "Latent space cartography: Visual analysis of vector space embeddings," *Comp. Graphics Forum*, pp. 67-78, 2019.
[38] H. Abdi, and L. J. Williams, "Principal component analysis," *Wiley Interdiscip. Rev. Comput. Stat.*, vol. 2, no. 4, pp. 433-459, 2010.
[39] F. Heimerl, and M. Gleicher, "Interactive analysis of word vector embeddings," *Comp. Graphics Forum*, vol. 37, no. 3, pp. 253-265, 2018.
[40] Z. Zheng, and L. Sun, "Disentangling latent space for vae by label relevant/irrelevant dimensions," in *CVPR*, pp. 12192-12201, 2019.
[41] P. Notin, J. M. Hernández-Lobato, and Y. Gal, "Improving black-box optimization in VAE latent space using decoder uncertainty," in *NIPS*, vol. 34, pp. 802-814, 2021.
[42] Z. Luo, A. Mishra, A. Achkar, J. Eichel, S. Li, and P. Jodoin, "Non-local Deep Features for Salient Object Detection," in *CVPR*, pp. 6593-6601, 2017.
[43] J. Hu, L. Shen, and G. Sun, "Squeeze-and-Excitation Networks," in *CVPR*, pp. 7132-7141, 2018.
[44] Z. Zheng, Y. Zhong, J. Wang, and A. Ma, "Foreground-Aware Relation Network for Geospatial Object Segmentation in High Spatial Resolution Remote Sensing Imagery," in *CVPR*, pp. 4095-4104, 2020.
[45] X. Ding, C. Shen, Z. Che, T. Zeng, and Y. Peng, "SCARF: A Semantic Constrained Attention Refinement Network for Semantic Segmentation," in *ICCVW*, pp. 3002-3011, 2021.
[46] H. Zhang, K. Dana, J. Shi, Z. Zhang, X. Wang, A. Tyagi, and A. Agrawal, "Context encoding for semantic segmentation," in *CVPR*, pp. 7151-7160, 2018.
[47] B. O. Ayinde, T. Inanc, and J. M. Zurada, "Redundant feature pruning for accelerated inference in deep neural networks," *Neural Networks*, vol. 118, pp. 148-158, 2019.
[48] K. Kahatapitiya, and R. Rodrigo, "Exploiting the redundancy in convolutional filters for parameter reduction," in *WACV*, pp. 1410-1420, 2021.
[49] M. Denil, B. Shakibi, L. Dinh, M. A. Ranzato, and N. J. A. i. n. i. p. s. De Freitas, "Predicting parameters in deep learning," in *NIPS*, vol. 26, 2013.
[50] F. Milletari, N. Navab, and S. Ahmadi, "V-Net: Fully convolutional neural networks for volumetric medical image segmentation," in *3DV*, pp. 565-571, 2016.
[51] S. Jadon, "A survey of loss functions for semantic segmentation," in *CIBCB*, pp. 1–7, 2020.
[52] H. Bogunović *et al.*, "RETOUCH: The Retinal OCT Fluid Detection and Segmentation Benchmark and Challenge," *IEEE Trans. Med. Imag.*, vol. 38, no. 8, pp. 1858-1874, 2019.
[53] N. Codella *et al.*, "Skin lesion analysis toward melanoma detection 2018: A challenge hosted by the international skin imaging collaboration (isic)," arXiv preprint arXiv:1902.03368, 2019.
[54] M. M. Fraz, P. Remagnino, A. Hoppe, B. Uyyanonvara, A. R. Rudnicka, C. G. Owen, and S. A. Barman, "An ensemble classification-based approach applied to retinal blood vessel segmentation," *IEEE Trans. Biomed. Eng.*, vol. 59, no. 9, pp. 2538-2548, 2012.
[55] D. Mahapatra, B. Bozorgtabar, and L. Shao, "Pathological retinal region segmentation from oct images using geometric relation-based augmentation," in *CVPR*, pp. 9611-9620, 2020.
[56] S. Reiß, C. Seibold, A. Freytag, E. Rodner, and R. Stiefelhagen, "Every annotation counts: Multi-label deep supervision for medical image segmentation," in *CVPR*, pp. 9532-9542, 2021.
[57] L.-C. Chen, Y. Zhu, G. Papandreou, F. Schroff, and H. Adam, "Encoder-Decoder with Atrous Separable Convolution for Semantic Image Segmentation," in *ECCV*, pp. 833-851, 2018
[58] O. Ronneberger, P. Fischer, and T. Brox, "U-Net: Convolutional networks for biomedical image segmentation," in *MICCAI*, pp. 234-241, 2015.
[59] O. Oktay *et al.*, "Attention U-Net: Learning Where to Look for the Pancreas," in *MIDL*, 2018.
[60] Z. Gu, J. Cheng, H. Fu, K. Zhou, H. Hao, Y. Zhao, T. Zhang, S. Gao, and J. Liu, "CE-Net: Context encoder network for 2D medical image segmentation," *IEEE Trans. Med. Imag.,* vol. 38, no. 10, pp. 2281-2292, 2019.
[61] F. Isensee, P. F. Jaeger, S. A. A. Kohl, J. Petersen, and K. H. Maier-Hein, "nnU-Net: a self-configuring method for deep learning-based biomedical image segmentation," *Nature Methods,* vol. 18, no. 2, pp. 203-211, 2021.
[62] S. Feng, H. Zhao, F. Shi, X. Cheng, M. Wang, Y. Ma, D. Xiang, W. Zhu, and X. Chen, "CPFNet: Context Pyramid Fusion Network for Medical Image Segmentation," *IEEE Trans. Med. Imag.,* vol. 39, no. 10, pp. 3008-3018, 2020.
[63] G. Xing, L. Chen, H. Wang, J. Zhang, D. Sun, F. Xu, J. Lei, and X. Xu, "Multi-Scale Pathological Fluid Segmentation in OCT With a Novel Curvature Loss in Convolutional Neural Network," *IEEE Trans. Med. Imag.,* vol. 41, no. 6, pp. 1547-1559, 2022.
[64] M. Wang, W. Zhu, F. Shi, J. Su, H. Chen, K. Yu, Y. Zhou, Y. Peng, Z. Chen, and X. Chen, "MsTGANet: Automatic Drusen Segmentation from Retinal OCT Images," *IEEE Trans. Med. Imag.,* vol. 41, no. 2, pp. 394-406, 2022.
[65] L. R. Dice, "Measures of the amount of ecologic association between species," *Ecology,* vol. 26, no. 3, pp. 297-302, 1945.
[66] K. H. Brodersen, C. S. Ong, K. E. Stephan, and J. M. Buhmann, "The Balanced Accuracy and Its Posterior Distribution," in *ICPR*, pp. 3121-3124, 2010.
[67] C. M. Seibold, S. Reiß, J. Kleesiek, and R. Stiefelhagen, "Reference-guided pseudo-label generation for medical semantic segmentation," in *AAAI*, pp. 2171-2179, 2022.
[68] K. Abhishek, G. Hamarneh, and M. S. Drew, "Illumination-based transformations improve skin lesion segmentation in dermoscopic images," in *CVPRW*, pp. 728-729, 2020.
[69] D. Li, J. Yang, K. Kreis, A. Torralba, and S. Fidler, "Semantic segmentation with generative models: Semi-supervised learning and strong out-of-domain generalization," in *CVPR*, pp. 8300-8311, 2021.
[70] R. Azad, A. Bozorgpour, M. Asadi-Aghbolaghi, D. Merhof, and S. Escalera, "Deep frequency re-calibration u-net for medical image segmentation," in *ICCV*, pp. 3274-3283, 2021.

[71] D. Dai, C. Dong, S. Xu, Q. Yan, Z. Li, C. Zhang, and N. Luo, "Ms RED: A novel multi-scale residual encoding and decoding network for skin lesion segmentation," *Med. Image Anal.,* vol. 75, pp. 102293, 2022.

[72] Z. Zheng, M. Oda, and K. Mori, "Graph Cuts Loss to Boost Model Accuracy and Generalizability for Medical Image Segmentation," in *ICCV*, pp. 3304-3313, 2021.

[73] S. Shit *et al.*, "clDice-a novel topology-preserving loss function for tubular structure segmentation," in *CVPR*, pp. 16560-16569, 2021.

[74] J. Deng, W. Dong, R. Socher, L. J. Li, L. Kai, and F.-F. Li, "ImageNet: A large-scale hierarchical image database," in *CVPR*, pp. 248-255, 2009.

[75] K. He, X. Zhang, S. Ren, and J. Sun, "Deep Residual Learning for Image Recognition," in *CVPR*, pp. 770-778, 2016.

[76] A. Paszke, S. Gross, F. Massa, A. Lerer, J. Bradbury, G. Chanan, T. Killeen, Z. Lin, N. Gimelshein, and L. Antiga, "Pytorch: An imperative style, high-performance deep learning library," in *NIPS*, pp. 8026-8037, 2019.

[77] R. Azad, M. Asadi-Aghbolaghi, M. Fathy, and S. Escalera, "Bi-directional ConvLSTM U-Net with densley connected convolutions," in *ICCVW*, pp. 0-0, 2019.

[78] M. Heidari, A. Kazerouni, M. Soltany, R. Azad, E. K. Aghdam, J. Cohen-Adad, and D. Merhof, "Hiformer: Hierarchical multi-scale representations using transformers for medical image segmentation," *arXiv preprint arXiv:2207.08518*, 2022.

[79] H. Wu, S. Chen, G. Chen, W. Wang, B. Lei, and Z. Wen, "FAT-Net: Feature adaptive transformers for automated skin lesion segmentation," *Med. Image Anal.,* vol. 76, pp. 102327, 2022.

[80] Y. Li, Y. Zhang, J.-Y. Liu, K. Wang, K. Zhang, G.-S. Zhang, X.-F. Liao, and G. Yang, "Global Transformer and Dual Local Attention Network via Deep-Shallow Hierarchical Feature Fusion for Retinal Vessel Segmentation," *IEEE Trans. Cybern.*, early access. doi: 10.1109/TCYB.2022.3194099, 2022.

[81] K. C. L. Wong, M. Moradi, H. Tang, and T. Syeda-Mahmood, "3D Segmentation with Exponential Logarithmic Loss for Highly Unbalanced Object Sizes," in *MICCAI*, pp. 612-619, 2018.

[82] R. Zhao, B. Qian, X. Zhang, Y. Li, R. Wei, Y. Liu, and Y. Pan, "Rethinking Dice Loss for Medical Image Segmentationin" in *ICDM*, pp. 851-860, 2020.

[83] C. H. Sudre, W. Li, T. Vercauteren, S. Ourselin, and M. Jorge Cardoso, "Generalised Dice Overlap as a Deep Learning Loss Function for Highly Unbalanced Segmentations," in *Deep Learn. Med. Image Anal. Int. Workshop Multimodal Learn. Clin. Decis. Support*, pp. 240-248, 2017.